\documentclass[english,amsmath,amssymb,showpacs,twocolumn]{revtex4}
\usepackage[T1]{fontenc}
\usepackage[latin1]{inputenc}
\usepackage[T1]{fontenc}
\usepackage[latin1]{inputenc}
\usepackage{babel}
\usepackage{graphics}
\begin{document}
\title{Quantum mechanical effect of path-polarization contextuality for a single photon}
\author{Alok Kumar Pan\footnote{apan@bosemain.boseinst.ac.in} and Dipankar Home\footnote{dhome@bosemain.boseinst.ac.in}}

\address{CAPSS, Department of Physics, Bose Institute, Sector-V, Salt Lake, Calcutta
700091, India}
\begin{abstract}
Using measurements pertaining to a suitable Mach-Zehnder(MZ) type setup, a curious quantum mechanical effect of contextuality between the path and the  polarization degrees of freedom of a polarized photon is demonstrated, without using any notion of realism or hidden variables - an effect that holds good for the  product as well as the entangled states. This form of experimental context-dependence is manifested in a way such that at \emph{either} of the two exit channels of the MZ setup used, the empirically verifiable \emph{subensemble} statistical properties obtained by an arbitrary polarization measurement depend upon the choice of a commuting(comeasurable) path observable, while this effect disappears for the \emph{whole ensemble} of photons emerging from the two exit channels of the MZ setup.
\end{abstract}
\maketitle
\section{Introduction}
Investigations of the implications of a possible `incompleteness' of quantum mechanics(QM) have resulted in striking discoveries of fundamental constraints which any realist model has to satisfy in order to be compatible with the empirically verifiable predictions of QM. One of such constraints is, of course, the comprehensively studied incompatibility between QM and the local realist models of quantum phenomena discovered using Bell's theorem\cite{bell64, aspect} and its variants\cite{hardy} - for a comparatively recent review of investigations in this area, see, for example, \cite{genovese}. The other constraint that, of late, has also been attracting an increasing attention is the one concerning the inconsistency between QM and the noncontextual realist(NCR) models(the Bell-Kochen-Specker theorem\cite{bell, kochen} and its variants[8-26]). It is this latter strand of study which leads to the present paper.  For this, we proceed by first recapitulating the essence of what is usually meant by the notion of `noncontextuality'.

Given any realist hidden variable model of quantum phenomena, the individual measured values of any dynamical variable are predetermined by the appropriate values of hidden variables($\lambda$'s) which are used in a realist model for a `complete specification' of the state of an individual quantum system. Now, the condition of `noncontextuality', in its most general form underlying its usual use, stipulates that the predetermined individual measured value of any dynamical variable, for a given $\lambda$, is the \emph{same} whatever be the way the relevant dynamical variable is measured. The question as to what extent this putative condition is compatible with the formalism of QM has been subjected to two different lines of study by exploring the implications of two separate facets of this condition. 

One of these is contingent upon assuming that the predetermined individual measured value of a given dynamical variable is independent of whatever be the choice of the other commuting(comeasurable) observable that is measured along with it. This condition has led to the formulation of a testable Bell-type inequality\cite{home} that is derived as a consequence of the NCR models, but is violated by QM for the entangled states by a finite amount, thereby enabling an empirical discrimination between QM and the NCR models\cite{simon, michler,hasegawa1}. Subsequently, the experimental investigation along this line has been enriched by further studies\cite{hasegawa2}. 

The other line of study on the issue of contextuality vis-a-vis QM is based upon a feature characterising the NCR models that can be expressed as follows:  For an individual measurement, the definite outcome obtained for an observable(say, $A$), as specified by a given hidden variable $\lambda$, be denoted by $v(A)$. Now, let $B$ be any other commuting(comeasurable)observable whose measured value in an individual measurement, as fixed by the \emph{same} given $\lambda$, be denoted by $v(B)$. Then, if one denotes an  individual outcome of a holistic measurement of the product observable $AB$ by $v(AB)$ which is determined by the \emph{same}  value of the hidden variable $\lambda$, the notion of noncontextuality is taken to imply the following condition(known as the `product rule') 
\begin{equation}
v(AB)=v(A)v(B)
\end{equation} 
which is assumed to hold good independent of the experimental procedure(context) used for measuring the joint observable $AB$ in a holistic way, and is also independent of the way the individual separate measurements of $A$  and $B$ are performed separately. This feature of noncontextuality was elegantly used by Mermin\cite{mermin} for formulating a proof of  quantum  incompatibility with the NCR models for two spin-1/2 particles that holds for \emph{any} state. Later, Cabello\cite{cabellosi} cast Mermin's proof\cite{mermin} in the form of a testable inequality involving the statistically measurable quantities - this inequality being violated by QM by a finite measurable amount for an \emph{arbitrary} two-qubit state. Subsequently, the state-independent quantum violation of this inequality has been experimentally corroborated using the polarization and the linear momentum degrees of freedom of  a single photon\cite{nature}.    

In contrast to the above two directions of study, in this paper we explore a third line of probing, initiated in a recent paper\cite{pan} that used a suitable path-spin entangled state of a spin-1/2 particle,  in which the issue of contextuality is probed \emph{within} the framework of QM, devoid of any reference to the NCR models. With a view of extending the ambit of this new line of study, the present paper reveals that that it is indeed possible to show a form of \emph{state-independent} contextuality \emph{within} QM for \emph{any}  state that is an entangled or a product state in the four dimensional space. By using photons and an appropriate setup we show that a statistically discernible effect of the path-polarization interdependence is manifested in terms of the operationally suitably defined \emph{subensemble} statistical properties of an arbitrary polarization measurement that depend upon the choice of  a  comeasurable(commuting)path observable. Interestingly, this context-dependence gets obliterated for the statistical results pertaining to the \emph{whole ensemble} of photons emerging from the setup used for the polarization measurement, whatever be the choice of the comeasurable path observable. Let us now proceed by first explaining the specifics of the setup(Figure 1) that is required for our demonstration. 
 
\section{The setup for preparing the required  product \emph{or} an entangled state}
In order to formulate our argument, the required path-polarization product state can be  prepared using a 50:50 beam splitter(BS1), while an entangled path-polarization state can be prepared by using a 50:50 BS1 in conjunction with a half wave plate(HWP) that is placed in one of the output ports of the BS1(Fig.1). The relevant path and the polarization measurements will be considered separately for these prepared states.

Let us consider that an ensemble of photons having horizontal polarized state $\left|H\right\rangle$ be incident on a  50:50 beam-splitter(BS1). Any given incident photon can then emerge along either the reflected  or the transmitted channel corresponding to the state designated by $\left| r\right\rangle$ or $\left|t\right\rangle$ respectively. Here note that for any given lossless beam-splitter, arguments using the unitarity condition show that for the photons incident on BS1, the phase shift between the transmitted and the reflected states of photons is essentially $\pi/2$\cite{zeilinger}. The prepared path-polarization product state after emerging from BS1 can then be written as 
\begin{eqnarray}
\label{pr}
\left|\Psi\right\rangle_{pr} =\frac{1}{\sqrt{2}}\left(\left|t\right\rangle 
+i  \left|r\right\rangle\right)\left|H\right\rangle
\end{eqnarray}

On the other hand, for preparing an entangled path-polarization state, photons in the channel corresponding to $|t\rangle$ are passed through a HWP that flips the polarization  $\left|H\right\rangle$ to $\left|V\right\rangle$. The state emerging from HWP can then be written as  
\begin{eqnarray}
\label{en}
\left|\Psi\right\rangle_{en} =\frac{1}{\sqrt{2}}\left(\left|t\right\rangle\left|V\right\rangle
+i  \left|r\right\rangle\left|H\right\rangle\right)
\end{eqnarray}
which is an entangled state between the path and the polarization degrees of freedom.

In writing both Eqs.(\ref{pr}) and (\ref{en}) we have taken into account a relative phase shift of $\pi/2$ between the states $\left|r\right\rangle$ and $\left|t\right\rangle$ that  arises because of the reflection from BS1. Eqs.(\ref{pr}) and (\ref{en}) represent the prepared states on which we will \emph{separately} consider the relevant  path and the polarization measurements for the purpose of showing the path-polarization interdependence for the product as well as for the entangled state. For this, we proceed as follows.
\vskip -0.5cm

\begin{figure}[h]
{\rotatebox{0}{\resizebox{12.0cm}{8.0cm}{\includegraphics{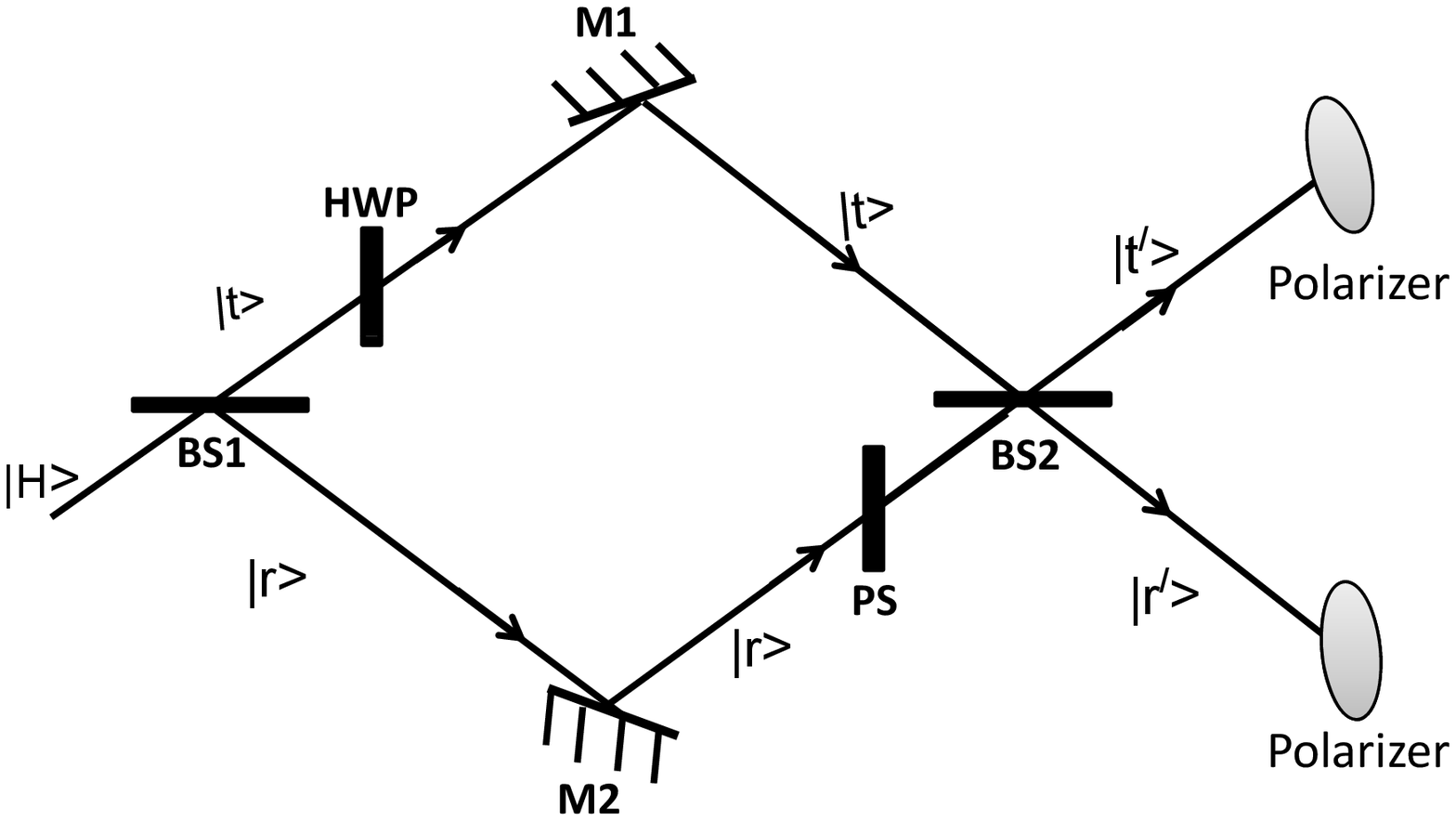}}}}
\vskip -2.0cm
\caption{\footnotesize Horizontally polarized (denoted by $\left|H\right\rangle$) photons enter the indicated Mach-Zehnder type setup through a 50:50 beam-splitter BS1, and  pass through the channels corresponding to  $|t\rangle$ and $|r\rangle$ thereby generating the path-polarization product state. For generating an entangled path-polarization state, a half wave plate (HWP) that flips the horizontally polarized state $\left|H\right\rangle$ into the vertically polarized state $\left|V\right\rangle$ is placed along one of the channels $|t\rangle$. Subsequently, for the required  measurements pertaining to the path observable, a phase-shifter(PS) is placed along the channel $\left|r\right\rangle$ that creates a relative phase shift between the channels  $\left|t\right\rangle$  and $\left|r\right\rangle$. The two channels are then recombined at a 50:50 beam-splitter BS2. The PS in conjunction with BS2  serves the purpose of path measurement(see the text). Finally, the measurement in an arbitrary polarization basis is made by using two polarizers that are placed along the two output channels $\left|t^{\prime}\right\rangle$ and $\left|r^{\prime}\right\rangle$ of the BS2. The path-polarization interdependence can then be demonstrated by considering the \emph{subensemble mean values} associated with each of the two output channels $\left|t^{\prime}\right\rangle$ and $\left|r^{\prime}\right\rangle$.}
\end{figure}

\section{The QM demonstration of path-polarization interdependence}
After passing through the mirrors M1 and M2, photons are subjected to a phase shifter(PS) along the channel $\left|r\right\rangle$ which introduces a relative phase shift of $\phi$ between the path channels $\left|r\right\rangle$ and $\left|t\right\rangle$. The two path states are then recombined at BS2, and the output path states $\left|t^{\prime}\right\rangle$ and $\left|r^{\prime}\right\rangle$ are respectively given by  
\begin{subequations}
\begin{eqnarray}
\label{psi34}
\left|t^{\prime}\right\rangle = \frac{1}{\sqrt{2}}\left(e^{i\phi}\left|r\right\rangle - i  \left|t\right\rangle\right)\\
\label{psi344}
\left|r^{\prime}\right\rangle = \frac{1}{\sqrt{2}}\left(-i e^{i\phi} \left|r\right\rangle + \left|t\right\rangle\right)
\end{eqnarray}
\end{subequations}

Eqs.(\ref{psi34},\ref{psi344}) show that, for a given linear combination of  $\left|t\right\rangle$ and $\left|r\right\rangle$, using the different values of  $\phi$, one can unitarily generate at the output of BS2 various linear combinations of $\left|t\right\rangle$ and $\left|r\right\rangle$ that correspond to different probability amplitudes of finding photons in the channels corresponding to $\left|t^{\prime}\right\rangle$ and $\left|r^{\prime}\right\rangle$. This, in turn, implies that PS in conjunction with BS2  can be regarded as corresponding to different choices of the path observables $\widehat {\beta}_{\phi}=\left|t^{\prime}\right\rangle \left\langle t^{\prime}\right| - \left|r^{\prime}\right\rangle \left\langle r^{\prime}\right|$ with eigenvalues $\pm 1$. Such observables,  in terms of actual measurements, correspond to different relative counts registered by the detectors placed along the channels represented by $\left|t^{\prime}\right\rangle$ and $\left|r^{\prime}\right\rangle$. 
Using Eqs.(\ref{psi34},\ref{psi344}) the path observable is of the form 
\begin{equation}
\beta_{\phi}= \left(\begin{array}{cl} 
    0& \ i e^{i\phi} \\  -i e^{-i\phi}  & \ 0 \end{array}\right) 
\end{equation} 
that can be written as the following linear combination of the Pauli matrices 
\begin{equation}
\beta_{\phi}= -sin\phi\ \widehat{\sigma}_{x}+ cos\phi\ \widehat{\sigma}_{y}= \vec{\sigma}.\vec{n}_{\phi}
\end{equation}
where $\vec{n}_{\phi}= -sin\phi \ \widehat{i}+ cos\phi\ \widehat{j}$. 

It is, therefore, evident from Eq.(5) that the path observable $\widehat{\beta}_{\phi}$ which is represented by the vector component given by Eq.(6) varies according to the magnitude  of the phase shift $\phi$, i.e., different choices of $\phi$ provide different contexts pertaining to the polarization measurements. 

Next, we consider the measurement of an arbitrarily chosen polarization variable, say, $\widehat \delta$ which is given by 
\begin{eqnarray}
\widehat{\delta}=\left|H^{\prime}\right\rangle \left\langle H^{\prime}\right| - \left|V^{\prime}\right\rangle \left\langle V^{\prime}\right|
\end{eqnarray}
whose eigenstates are  $\left|H^{\prime}\right\rangle= cos \alpha \left|H\right\rangle+ sin \alpha \left|V\right\rangle$ and $\left|V^{\prime}\right\rangle= sin\alpha\left|H\right\rangle - cos \alpha\left|V\right\rangle$, where $\alpha$ denotes the orientation of the two polarizers that are placed separately along the channels $\left|t^{\prime}\right\rangle$ and $\left|r^{\prime}\right\rangle$.

Now, we will consider the expectation value of the polarization observable $\widehat{\delta}$ that involves contributions from both the output subensembles  corresponding to polarizers \emph{separately} placed along the channels $\left|t^{\prime}\right\rangle$ and $\left|r^{\prime}\right\rangle$. The \emph{subensemble mean values} of $\widehat\delta$ measured in each of the two output channels, calculated using either the prepared product or an entangled path-polarization state given by Eqs.(2) and (3) respectively, are denoted by $(\bar{\delta})_{t^{\prime}}$ and $(\bar{\delta})_{r^{\prime}}$, whence 
\begin{equation}
\label{meansub}
\left\langle \widehat{\delta}\right\rangle_{\Psi_{pr/en}}=(\bar{\delta})_{t^{\prime}}+ (\bar{\delta})_{r^{\prime}}
\end{equation} 
where the subscript $\Psi_{pr/en}$ represents the prepared entangled or product states given by Eq.(2) or (3) respectively.

Note that all the three quantities occurring in the equality given by Eq.(\ref{meansub}) have the same operational status as far as their \emph{statistical reproducibility} is concerned. But there is a crucial distinction between the left and the right hand sides of Eq.(\ref{meansub}) with respect to the issue of path-polarization  interdependence. The quantity on the left  hand side of Eq.(\ref{meansub}), the expectation value of $\langle\widehat{\delta}\rangle_{\Psi_{pr/en}}$ pertaining to the whole ensemble, is \emph{independent} of which path observable is measured along with it. On the other hand,  each of the quantities on the right hand side of Eq.(\ref{meansub}), the subensemble mean values  denoted by $(\bar{\delta})_{t^{\prime}} $ and $(\bar{\delta})_{r^{\prime}} $  are contingent upon the choice of the comeasurable path observable.  This can be seen by considering the polarization measurement outcomes relevant to any one of the two output subensembles. 

In order to display the manifestation of this form of context-dependence within QM,  we consider two different experiments involving measurements of the  path observable $\widehat{\beta}_{\phi}$ and the polarization variable $\widehat{\delta}$, first for the prepared product state given by Eq.(2), and then for the prepared entangled state given by Eq.(3).
\subsection{Path-polarization context-dependence for a product state}
We first consider the path-polarization product state given by Eq.(2) as the input state for which the state that emerges from BS2 can be written as 
\begin{eqnarray}
\left|\Phi\right\rangle_{pr}= \frac{1}{2}\left[i\left|t^{\prime}\right\rangle\left(1+e^{i\phi}\right)
+   \left|r^{\prime}\right\rangle\left(1- e^{i\phi}\right)\right]\left|H\right\rangle
\end{eqnarray}

Then it follows that the expectation value of the polarization observable $\widehat\delta$  pertaining to the \emph{whole ensemble} of photons emerging from BS2 is of the form

\begin{equation}
\left\langle{\widehat{\delta}}\right\rangle_{\Psi_{pr}}= cos 2\alpha
\end{equation}

which comprises the respective subensemble polarization mean values calculated from Eq.(9) given by
\begin{eqnarray}
(\bar{\delta})_{t^{\prime}}=\frac{(1+cos \phi)cos 2\alpha}{2}; \hskip 0.2cm (\bar{\delta})_{r^{\prime}}= \frac{(1 - cos \phi)cos 2\alpha}{2}
\end{eqnarray}

Next, we come to the crux of our argument indicated as follows that hinges on two different choices of $\phi$, and where the superscript ${\beta}_{0}$(${\beta}_{\pi/2}$) is used to denote the choice of the path observable specifying the given context: 

$(a)$ Taking $\phi=0$,  this  implies the  choice of a particular  path observable $\widehat{\beta}_{0}=\vec{\sigma}.\vec{n}_{0}$ where $\vec{n}_{0}=\widehat{j}$. In this case, using Eq.(11), we obtain 

\begin{eqnarray}
(\bar{\delta})^{({\beta}_{0})}_{t^{\prime}}= cos 2\alpha; \hskip 0.3cm 
(\bar{\delta})^{({\beta}_{0})}_{r^{\prime}}= 0
\end{eqnarray}

while, using Eq.(2), the polarization expectation value for the whole ensemble, $\langle\widehat{\delta}\rangle_{\Psi_{pr}}=cos 2\alpha$.

$(b)$ Taking $\phi=\pi/2$, this implies the choice of a different path observable $\widehat{\beta}_{\pi}=\vec{\sigma}.\vec{n}_{\pi/2}$ where $\vec{n}_{\pi/2}= -\widehat{i}$.  Consequently, using Eq.(11), we obtain

\begin{eqnarray}
(\bar{\delta})^{({\beta}_{\pi/2})}_{t^{\prime}}= \frac{cos 2\alpha}{2}; \hskip 0.3cm (\bar{\delta})^{({\beta}_{\pi/2})}_{r^{\prime}}=  \frac{cos 2\alpha}{2}
\end{eqnarray}
while, using Eq.(2), the polarization expectation value for the whole ensemble remains the \emph{same}, $\langle\widehat{\delta}\rangle_{\Psi_{pr}}=cos 2\alpha$. 

It is then evident from Eqs.(10-13) that, while the quantum expectation value $\langle\widehat{\delta}\rangle_{\Psi_{pr}}$ of the observable $\widehat\delta$ pertaining to the \emph{whole ensemble} remains the \emph{same} for both the choices of $\widehat{\beta}_{0}$ and $\widehat{\beta}_{\pi/2}$, the path-polarization context-dependence gets manifested in terms of the \emph{subensemble polarization mean values} given by the  testable quantities   $(\bar{\delta})^{({\beta}_{0}, {\beta}_{\pi/2})}_{t^{\prime}}$ and $(\bar{\delta})^{({\beta}_{0}, {\beta}_{\pi/2})}_{r^{\prime}}$. To put it  precisely, in our example, the interdependence between the path and the polarization degrees of freedom has the following operational meaning
\begin{eqnarray}
(\bar{\delta})^{({\beta}_{0})}_{t^{\prime}}\neq(\bar{\delta})^{({\beta}_{\pi/2})}_{t^{\prime}} ; \hskip 0.3cm (\bar{\delta})^{({\beta}_{0})}_{r^{\prime}}\neq(\bar{\delta})^{({\beta}_{\pi/2})}_{r^{\prime}}
\end{eqnarray}

 i.e., the subensemble mean value of the polarization  variable $\widehat{\delta}$ depends upon \emph{which} of the path observables $\widehat{\beta}_{0}$ or $\widehat{\beta}_{\pi/2}$ is comeasured, where both $\widehat{\beta}_{0}$ and $\widehat{\beta}_{\pi/2}$ commute with $\widehat{\delta}$. 

\subsection{Path-polarization context-dependence for an entangled state}
Now, we consider the path-polarization entangled state given by Eq.(3) as the input state for which the state that emerges from BS2 can be written as 
 
\begin{eqnarray}
\left|\Phi\right\rangle_{en}= \frac{1}{2}\left[i\left|t^{\prime}\right\rangle\left( \left|V\right\rangle + e^{i\phi}\left|H\right\rangle\right)
+  \left|r^{\prime}\right\rangle\left(\left|V\right\rangle - e^{i\phi}\left|H\right\rangle\right)\right]
\end{eqnarray}

It follows from Eq.(15) that corresponding to the prepared path-polarization entangled state $\left|\Psi\right\rangle_{en}$ given by Eq.(3), the expectation value of the polarization variable $\widehat\delta$  for the \emph{whole ensemble} of photons emerging from the beam-splitter BS2 is given by 

\begin{equation}
\langle\widehat{\delta}\rangle_{\Psi_{en}}=0
\end{equation}
which is made up of the respective subensemble polarization mean values calculated from Eq.(15), which are of the form 

\begin{eqnarray}
(\bar{\delta})_{t^{\prime}}=\frac{ sin 2\alpha \ cos\phi}{2}; \hskip 0.3cm (\bar{\delta})_{r^{\prime}}= - \frac{ sin 2\alpha \ cos\phi}{2}
\end{eqnarray}

Then, similar to the argument given above for the input product state, in this case too Eq.(14) holds good, i.e., 
the path-polarization context-dependence gets manifested in terms of the \emph{subensemble polarization mean values} given by the  testable quantities   $(\bar{\delta})^{({\beta}_{0}, {\beta}_{\pi/2})}_{t^{\prime}}$ and $(\bar{\delta})^{({\beta}_{0}, {\beta}_{\pi/2})}_{t^{\prime}}$,  while the quantum expectation value  $\langle\widehat{\delta}\rangle_{\Psi_{en}}$ pertaining to the \emph{whole ensemble} remains the \emph{same} for both the choices of $\widehat{\beta}_{0}$ and $\widehat{\beta}_{\pi/2}$.

\section{The significance and outlook}
The essence of what is demonstrated in this paper is as follows. A statistically discernible signature of interdependence between the path and the polarization degrees of freedom of polarized photons is revealed - an effect which is quantum mechanically calculable in terms of the measured values of a polarization variable pertaining to the operationally well-defined  \emph{subensembles} that comprise the final output ensemble at the two exit channels of our setup. The subensemble polarization mean value registered at either of the two exit channels \emph{varies} according to the choice of the comeasurable(commuting) path observable(whose choice is fixed by the magnitude of the phase shift that is introduced by PS in the channel $\left|r\right\rangle$) But, importantly, such a variation takes place by \emph{preserving} the context-independence of the polarization expectation value that is defined for the whole output ensemble. 

In other words, we show that for an arbitrarily prepared state that can be either a path-polarization entangled or a product state, a form of `parameter dependence' is displayed in this example in a way that is  restricted to the \emph{subensemble statistics}. This is quite distinct from the issue of `parameter independence'\cite{shimony} for the EPR-Bohm type states involving the polarization variables of  two spatially separated photons where the `parameter independence' is in the sense that \emph{no} statistically discernible effect in any form can be detected in any one of the two wings of the EPR-Bohm pair that depends upon the measurement setting in the other wing - a possibility which is ruled out by the much-discussed `no-signaling' condition\cite{bohmbook1}. But, in contrast, in our example, for an arbitrary input state considered in our example for a polarized photon, there is no such constraint which forbids the statistical manifestation of an intraparticle path-polarization context-dependence.  While such an effect does occur in our example, curiously, it is confined to the subensemble statistics in such a way that the `path-polarization interdependence' disappears for the statistics of the whole ensemble. 

In terms of the two distinct aspects of the NCR models as discussed  earlier, we may recall that the statistically verifiable inequalities for the entangled as well as the product states have been analyzed, both theoretically and experimentally, thereby highlighting an incompatibility between QM and the NCR models.  However, it needs to be stressed that these earlier demonstrations necessarily involve the assumption of the notion of realism that is used in tandem with the feature of noncontextuality at the level of individual measurement outcomes. In contrast, the statistically verifiable manifestation of the quantum mechanical `path-polarization interdependence' shown in this paper is entirely \emph{independent} of any notion of `realism' or `hidden-variables', because this effect is demonstrated essentially within the ambit of QM and, crucially, is independent of the nature of the input state which can be either a path-polarization product or an entangled state. A significant point to stress here is that a similar effect of interdependence showing a nonlocal connection between the entangled variables of the spatially separated photons cannot be demonstrated within QM unless one takes recourse to the notion of `realism' or `hidden-variables'. This fundamental distinction between quantum nonlocality and contextuality brought out by the example analyzed in this paper as well as in an earlier work\cite{pan} call for careful probing. Such investigations may provide useful insights into the type of constraints that would restrict the nonlocal realist models in the light of the recent studies\cite{gisin, gro, lepu} based on Leggett's work\cite{leggett}. 

\section*{Acknowledgements} Authors thanks H. Rauch and Y. Hasegawa for the useful interactions that served as a prelude to this work.  AKP recalls the discussions, in particular, with A. Cabello during his visit to Benasque Center for Science, Spain. AKP acknowledges the Research Associateship of Bose Institute, Kolkata. DH acknowledges the project funding from DST, Govt. of India and  support from Center for Science and Consciousness, Kolkata. 

\end{document}